\documentclass{aa501}
\usepackage{graphicx}
\begin{document}

%
   \title{On the orbital period of the cataclysmic
variable \object{RZ\,Leonis}}


\author{R.E.\ Mennickent \inst{1}\fnmsep\thanks{Based on observations
obtained at the European Southern Observatory, ESO proposal 54.E-0812}
\and  Claus Tappert \inst{1,2}.}

   \institute{Dpto. de Fisica, Facultad de Ciencias Fisicas y Matematicas,
Universidad de Concepcion, Casilla 160-C, Concepcion, Chile\\ email:
rmennick@stars.cfm.udec.cl
\and Dipartimento di Astronomia, Universita di
Padova,  Vicolo dell'Osservatorio 2, I-35122 Padova, Italy \\
e-mail: ctappert@pd.astro.it}

   \offprints{R.E.\ Mennickent}
   \mail{rmennick@stars.cfm.udec.cl}
   \date{}

   \abstract{
We present a time-resolved study of the Balmer emission lines of
\object{RZ Leo}. From the analysis of the radial velocities  we
find an orbital period of  0.07651(26) d.
 \keywords{
   Stars: individual: \object{RZ Leonis},
   Stars: novae, cataclysmic variables,
   Stars: fundamental parameters,
   Stars: evolution,
   binaries: general }}

   \maketitle

\section{Introduction}
 
The cataclysmic variable  RZ Leo has been recently confirmed as an SU UMa
type dwarf nova after the discovery of superhumps during the December 2000
superoutburst (Ishioka et al.\ 2000). The photometric
period detected in quiescence by Mennickent et al.\ (1999),
$P_{\rm ph}$ = 0.0756(12) d,
was also reported during early stages of the superoutburst
and thus possibly represents the orbital period of the binary.
In this paper we present time resolved spectra of RZ Leo. From the study of the
radial velocities of the H$\alpha$  emission line we found a
spectroscopic period of $P_{\rm sp}$ = 0.07651(26) d.
We furthermore investigate the variation of the quiescent emission line
profile, and find evidence for a prominent emission component from the
hot spot region.

\section{Observations and data reduction}

A total of 46 spectra with individual exposure times of 10 minutes were obtained
during two observing runs at La Silla and
Las Campanas observatories. The observing schedule and instrumental setup are shown in
Table 1. All CCD frames were processed in the standard way,  using IRAF (distributed by the National Optical
Astronomy Observatories).
One-dimensional spectra were extracted, sky-subtracted and wavelength calibrated.

  \begin{table}
 \caption[]{Table 1. Journal of Observations. Setup $A$ corresponds to ESO
observations with the 2.2 m telescope using EFOSC2 and grating 4. Setup $B$
corresponds to LCO observations with the 2.5 m telescope using the Modular
Spectrograph and grating 6.}
\begin{center}
\begin{tabular}{|c|c|c|c|}
\hline \hline \multicolumn{1}{c}{Date (UT)}&
\multicolumn{1}{c}{HJD start} &
\multicolumn{1}{c}{$n_{\rm spec}$}&
\multicolumn{1}{c}{setup/resolution}  \\ \hline
 10/02/95& 244\,9758.7706 &14   &A/6\AA \\
11/02/95& 244\,9759.7276 &15   &A/6\AA \\
21/03/95& 244\,9797.7109 &7    &B/4.5\AA\\
23/03/95& 244\,9799.7113 &10   &B/4.5\AA \\
\hline  \noalign{\smallskip}   \hline  \end{tabular}  \end{center}
\end{table}

\section{Spectrophotometric measurements and radial velocity study}

   \begin{table}
 \caption[]{Spectroscopic measurements of the averaged spectra.
 $W_{\lambda}$ and $\Delta \lambda$ are in \AA, $FWZI$ and $FWHM$ in
 km/s.}
\begin{center}
\begin{tabular}{|c|c|c|c|c|}
\hline \hline
\multicolumn{1}{c}{$Line$ } &
\multicolumn{1}{c}{$W_{\lambda}$}&
\multicolumn{1}{c}{$\Delta \lambda$}&
\multicolumn{1}{c}{$FWZI$ }&
\multicolumn{1}{c}{$FWHM$} \\ \hline
 H$\alpha$ & -130 & 600 & 4300 & 2130 \\
H$\beta$ & -14 & 1150 & 3600 & 2610 \\
\ion{He}{I}\,5875 &-34 & 1250 & 3475 & 2170 \\
\hline  \noalign{\smallskip}   \hline  \end{tabular}  \end{center}
\end{table}

\noindent  The system was found in quiescence. The spectrum presented
here is characterized  by double emission lines of H$\beta$,
\ion{He}{I}\,5875 and H$\alpha$ (Fig.\,1). In Table 2 we give mean
spectroscopic parameters for the main emission lines: equivalent widths,
half-peak separation $\Delta \lambda$, full width at half maximum $FWHM$
and full width at zero intensity $FWZI$.\\


\noindent Radial velocities
were measured using a single Gaussian fit to the line profile. These values
were analysed using both the Scargle and the AOV algorithm (Scargle 1982;
Schwarzenberg-Czerny 1989; resp.) implemented in MIDAS. The results in
Fig.\,2 show  several possible
frequencies with similar power around a  peak frecuency of about 13
cycles/d. In order to discriminate between the possible frequencies and
derive the true period, we fitted the data with a sine function
corresponding to the peak frecuency. Then a Monte Carlo simulation was
applied in such a way that to each data point a random value from an
interval consisting of $\pm$ the sigma of the sine fit was added. The
resulting data set was again analysed with the Scargle algorithm, and the
highest peak was registered. After a thousand repetitions, the average
value within a small interval (0.2 cycles/d, determined by visual
inspection of the periodogram) around the more recurrent period and its
sigma were taken as the resulting period and its error, respectively. The
histogram in Fig.\,3 shows the resulting distribution for a binsize of
0.01 cycles/d.
The derived period  was

\begin{displaymath}  P_{sp} = 0.07651(26)~{\rm d} =
1.8363(62)~{\rm h}. \end{displaymath}

\noindent which is in excellent agreement with the photometric value $P_{\rm ph}$ =
0.0756(12) d found by Mennickent et al. (1999).
This value places RZ Leo in the midst of the SU UMa period distribution,
and near the location of the well-known systems \object{Z Cha} and
\object{SU UMa}.   From the sine fit to the radial velocity data set
(Fig.\, 4) we furthermore obtain a semi-amplitude $K_1 = 91 \pm 12$ km
s$^{-1}$ and a time HJD$_0$ = 2449759.7489 $\pm$ 0.0022 corresponding to
the superior conjunction of the emitting source (positive to negative
crossing of the radial velocity curve). Note that these values were
derived by measuring the whole line profile, and therefore cannot be
identified a priori with the corresponding parameters of the white dwarf.

The February data set also includes accompanying photometric measurements
in the V passband. A nearby comparison star  was placed in the slit
at the same time that the variable, and  differential slit-magnitudes
were extracted from the spectra counts. A few V images taken during the
night permitted to roughly calibrate these magnitudes  using nearby
comparison stars whose  standard magnitudes were taken from Mennickent
et al. (1999). We estimate that the zero point for the $V$ magnitudes so
obtained is accurate to at least 0.1 mag. Results are discussed in the
next Section.

    \begin{figure}
  \scalebox{.3}[.3]{\includegraphics{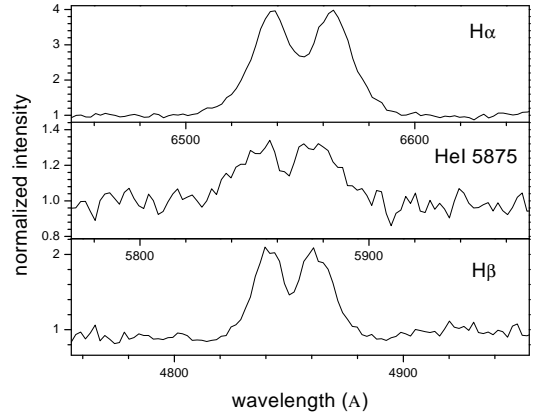}}
  \caption{Averaged emission lines during February 1995.}  \label{spectrum}
\end{figure}

\begin{figure}
  \includegraphics[angle=270,width=8cm]{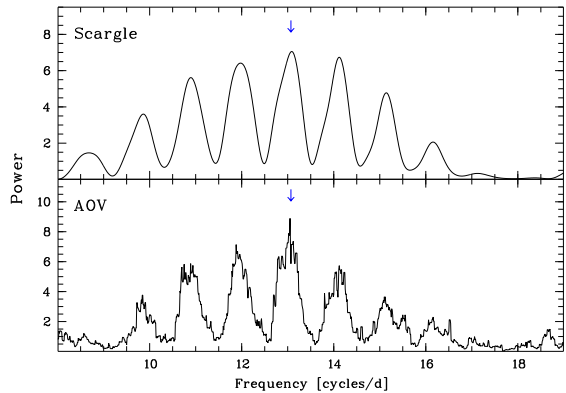}
  \caption{Scargle (top) and AOV (bottom) periodogram of the
radial velocity data. The arrows indicate the frequency corresponding to the
derived period.}
 \label{periodogram}
 \end{figure}

  \begin{figure}
  \includegraphics[angle=270,width=8cm]{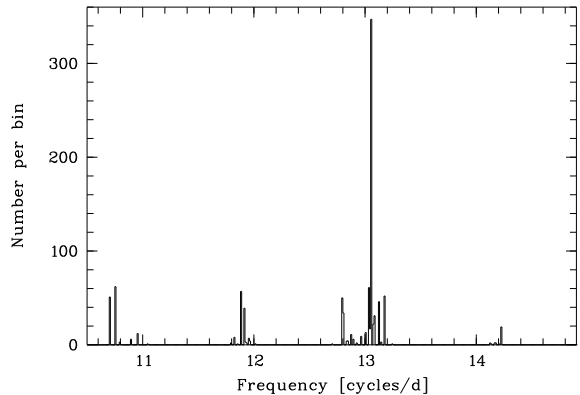}
  \caption{Histogram of the maximum-peak frequency.}
 \label{periodogram}
 \end{figure}

\section{Discussion}

\subsection{On the orbital and superhump period}

Within the errors, the derived spectroscopic period corresponds well to the
photometric period in quiescence, and thus very likely is the orbital period of
the binary. A similar period was also detected photometrically during the early
stages of the superoutburst. This supports the current understanding of the
formation of ``early superhumps'', which still resemble the orbital period,
and which later evolve into ``full superhumps'' as the accretion disc is
deformed and starts to precess. Considering the superhump period of
0.07857(22)d (Ishioka et al.\ 2000) we find a period excess between the
superhump period $P_{\rm SH}$ and the orbital period $P_{\rm orb}$ of
\begin{displaymath}
 \epsilon = \frac{P_{\rm SH}-P_{\rm orb}}{P_{\rm orb}} = 0.0269(81),
\end{displaymath}
which is a typical value for an SU UMa star of this period (e.g.,
Patterson 1998).

   \begin{figure}
  \includegraphics[angle=0,width=8cm]{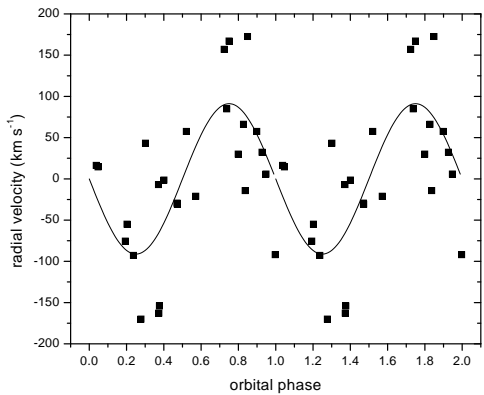}
  \caption{The H$\alpha$ emission line radial velocity versus the
orbital  phase and the best sine fit.}
\label{periodogram}  \end{figure}

\subsection{On the changing nature of the hotspot}


  \begin{figure}
  \includegraphics[angle=0,width=9cm]{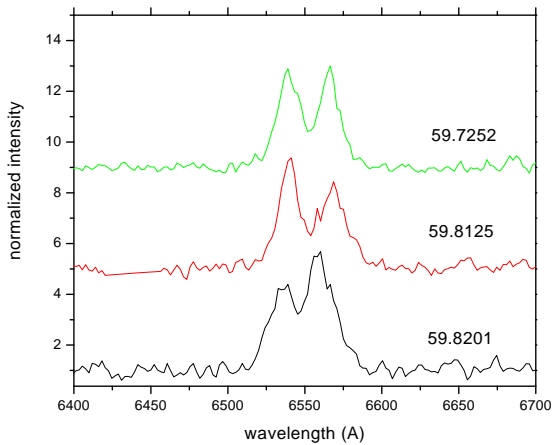}
  \caption{Selected  H$\alpha$ profiles and their $HJD$
  (minus 244\,9700) showing $V/R$
variations. A vertical spacing of 3 units have been applied to the
profiles.}
\label{periodogram}  \end{figure}

As in other well known dwarf novae, variations
in the relative strength of the violet and red peak intensity are
observed  (Fig.\, 5).  These V/R variations could be attributed to the
changing aspects of a hotspot in the  disk-stream interacting region. This
hotspot seems to be also the origin for the variations of the continuum
and line strength observed in Fig.\,6. The fact that the photometric
maximum corresponds to the equivalent width maximum indicates that the
hotspot is the source of both continuum and line extra emission. In
addition,  the displacement of the maximum phase at different
epochs is notable. This could indicate displacements of the hotspot in the binary
frame of rest, which was also suggested by Mennickent et al.\,(1999) to
explain the long-term photometric variations. The position of the hotspot
is that expected for the stream-disk interacting region in February, but
not in March. Anomalous hotspot positions have been also observed by the
method of Doppler Tomography and they are not yet fully understood
(Szkody 1992).

  \begin{figure}
  \includegraphics[angle=0,width=9cm]{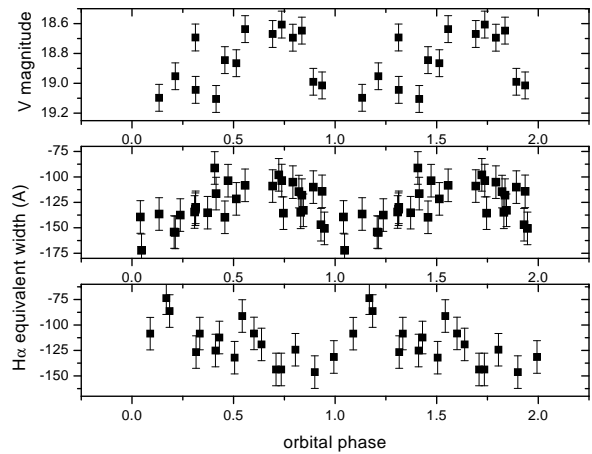}
  \caption{The $V$ magnitude for the first night of February (upper
panel). The H$\alpha$ emission line equivalent
widths for February (middle panel) and March
(below panel). A small shift was applied to the equivalent width of
February to correct for different nightly mean.
Note the non-coherent nature of the equivalent width oscillations.}
\label{periodogram} \end{figure}

\section{Conclusions}

   \begin{enumerate}
      \item We have found that the orbital period of \object{RZ\,Leonis} is
0.07651(26)
\item The hotspot seems to be the source of both continuum and the
additional line emission
\item The position of the hotspot in the system of rest of the binary seems
to be variable    \end{enumerate}

\begin{acknowledgements}
This work was supported by Grant Fondecyt 1000324 and DI 99.11.28-1.
 
\end{acknowledgements}

 \end{document}